\documentclass[preprint,12pt]{elsarticle}

\usepackage{amssymb}
\usepackage{amsmath}
\usepackage{float}
\usepackage{subcaption}
\usepackage{graphicx}
\usepackage{booktabs}
\usepackage{array}
\usepackage{siunitx}
\usepackage{booktabs}
\usepackage{array}
\usepackage{makecell}
\usepackage{caption}

\journal{Journal of Computational Physics}

\begin{document}

\begin{frontmatter}



\title{Volume penalization method for simulating flows around a rotating solid with multiple reference frame and sliding mesh}


\author{Ming Liu} 
\author{Yosuke Hasegawa\corref{cor1}} 
\cortext[cor1]{Corresponding author.}

\affiliation{organization={Center for Research on Innovative Simulation Software, Institute of Industrial Science, The University of Tokyo},
            addressline={4-6-1 Komaba, Meguro-ku}, 
            city={Tokyo},
            postcode={153-8505}, 
            state={},
            country={Japan}}

\begin{abstract}
Despite the significant role of turbomachinery in fluid-based energy transfer, precise simulation of rotating solid objects with complex geometry is a challenging task. In the present study, the volume penalization method (VPM) is combined with multiple reference frame (MRF) and sliding mesh (SLM), respectively, so as to develop immersed-boundary approaches for simulating flows around a rotating solid. The level-set function is adopted to represent arbitrary geometries embedded in Cartesian grids. The VPM body-forcing terms in the momentum equation are proposed for MRF and SLM, respectively, so as to build unified governing equations for both fluid and solid regions. The flows around a rotating cuboid under various rotating speeds are simulated by the present schemes, namely, VPM with MRF, and VPM with SLM, and compared to corresponding simulations by the body-fitted method (BFM). The results suggest the relative deviations of predicted pressure drop and torque between the present VPM and BFM are around 5\%, demonstrating the validity of the present VPM.
\end{abstract}

\begin{keyword}
volume penalization method, multiple reference frame, sliding mesh, turbomachinery
\end{keyword}

\end{frontmatter}



\section{Introduction}
\label{sec1}

Turbomachinery plays a significant role in various industrial applications to transfer energy between a rotor and a fluid, e.g., turbines for power generation, and pumps for providing energy to fluids. Electricity consumption of pumps and fans accounts for 30\%-35\% of global energy consumption \cite{Yao24}. This highlights its significant potential for energy savings through pump performance optimization, which is crucial for achieving a carbon-neutral society. Although significant progress has been made in the field of computational fluid dynamics, simulating flows past turbomachinery remains challenging due to the presence of moving boundaries. In order to take the interaction between rotating objects and surrounding stationary solid boundaries into consideration, the entire domain is often divided into static and rotating regions. Depending on the way to treat the rotating region, existing numerical approaches are classified into multiple reference frame (MRF) \cite{Romero14}\cite{Alonso23} and sliding mesh (SLM) \cite{Kang14}\cite{Zhang15} techniques.

MRF simplifies the simulation by fixing a rotating object inside the rotating region, while only inertia terms arising from the rotation of the frame of reference are taken into account in the momentum equation \cite{Mangani16}-\cite{Peng19}. Since rotating objects are virtually fixed in static grids, the difficulties in dealing with moving boundaries are removed. However, it neglects the unsteady interaction between the rotating solid objects and the external stationary boundaries, and therefore it often fails to capture detailed transient characteristics.

In order to reproduce the unsteady nature of a rotating object, the body forcing method, typified by the actuator disk approach \cite{Calaf10}-\cite{Jia24} has been proposed. The effects of a rotating object on the surrounding flow field are modelled by its total thrust and introduced as a body force into the governing equation. This method is efficient in analyzing wake dynamics and is widely applied in simulations of wind turbines in large wind farms \cite{Jia24}. However, since the effects on a rotating object are modelled by a diffused body force, it may fail to reproduce the detailed dynamics near the boundary between the rotating object and the surrounding fluid. There also exists another type of body forcing method developed under the immersed boundary framework, namely, the direct forcing method \cite{Goldstein93}, where an additional forcing term is added to the momentum equation so as to enforce the boundary condition on the solid surface \cite{Zhong09}-\cite{Du16}. At each time step, it is required to map a rotating object to stationary grids, making it inefficient to deal with complex geometries \cite{Chen21}-\cite{Specklin18}.

SLM is widely adopted to reproduce detailed transient flow fields around a rotating solid \cite{Steijl08}-\cite{Liang20}. It explicitly takes the rotation of a solid object into account by rotating the grids inside the rotating region at a predefined rotation speed \cite{Rai86}\cite{Verzicco23}. This makes the rotating object stationary in the rotating frame. In the existing SLM simulations, body-fitted meshes have often been employed. Although sufficient accuracy can be obtained, it always encounters difficulties in generating high-quality body-fitted meshes that are mostly curvilinear and non-orthogonal for turbomachinery \cite{Mittal25}.

To solve the problem of generating body-fitted mesh, the immersed boundary method (IBM) has attracted much attention recently \cite{Mittal25}, which embeds an arbitrary solid geometry on Cartesian grids. Generally, IBM can be categorized into sharp and diffused approaches, depending on whether the fluid-solid interface is kept sharp or diffused within a few cells \cite{Schneider05}. The volume penalization method (VPM) \cite{Kadoch12}\cite{Kumar24}, a diffused-interface IBM, is originally motivated by modelling solid walls as porous media with nearly zero permeability. By introducing artificial body force terms in the momentum equation, it forces the local velocity inside the solid region to the target velocity with a unified governing equation for both fluid and solid regions. The VPM has been extended to various flow and heat transfer problems \cite{Liu23}\cite{Kametani20} and shows great potential for the combination of adjoint-based optimization. Recent studies \cite{Chen24}-\cite{Liu25} have successfully achieved drag reduction and heat transfer enhancement in different thermofluidic systems. However, the VPM for a rotating solid object has not been established, and it is a key step to realize shape/topology optimization of turbomachinery.

In the present study, novel VPMs to simulate fluid flow around a solid object are developed in the framework of MRF and SLM, respectively. The following sections are organized as follows. The numerical methodologies based on VPM and combined with MRF and SLM are presented in Sec. \ref{sec2}. In Sec. \ref{sec3}, the proposed scheme is validated through simulations of flows past a rotating cuboid by comparing with body-fitted simulation results. Finally, the conclusions of the present study are summarized in Sec. \ref{sec4}.

\section{Numerical methodology}
\label{sec2}
\subsection{Problem description}
\label{subsec2_1}
\begin{figure}[H]
\centering
\includegraphics[width=0.8\textwidth]{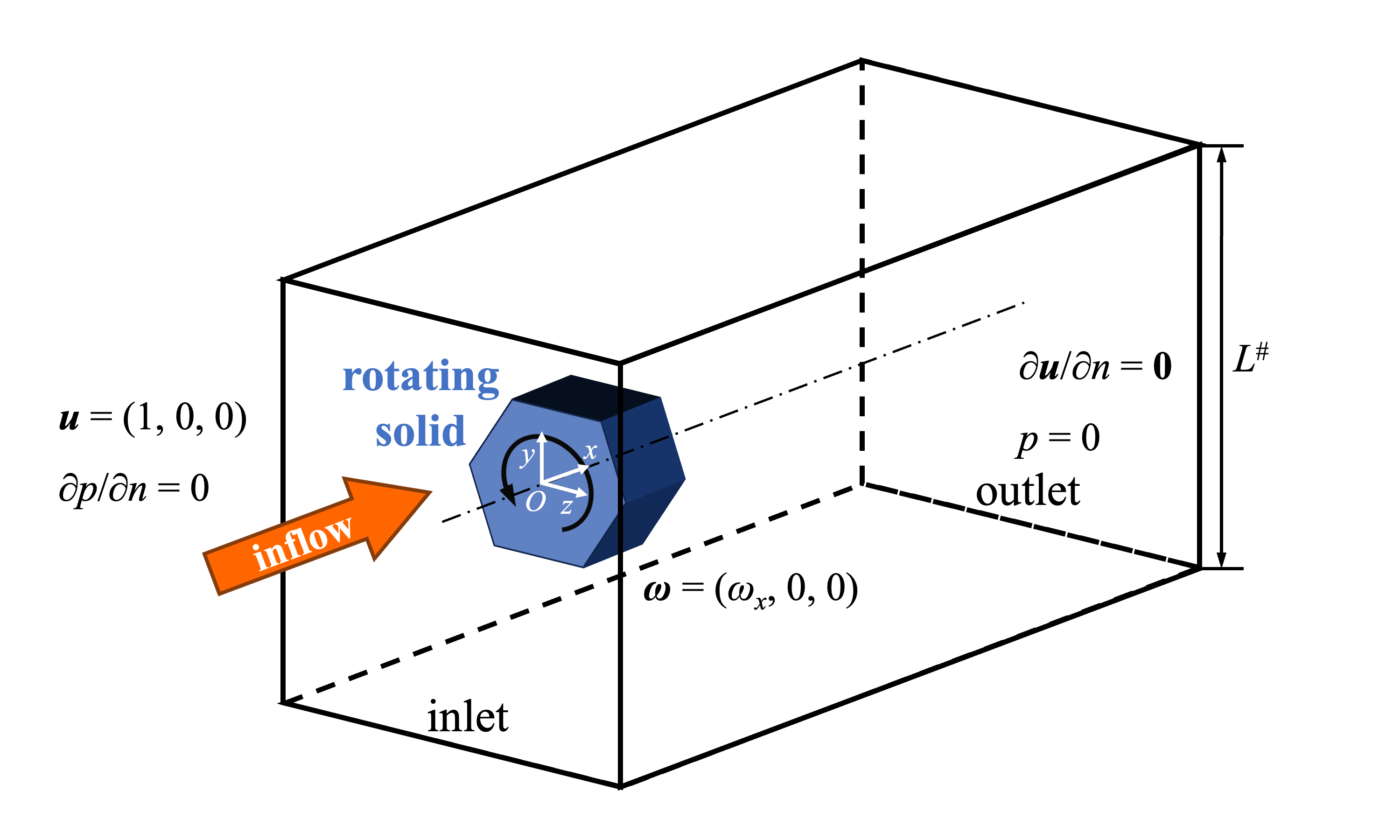}
\caption{Schematic for the flow around rotating solids.}\label{fig1}
\end{figure}

An incompressible laminar flow past a rotating solid object is investigated in the present study as illustrated in Fig. \ref{fig1}. The streamwise direction is defined as the $\mathit{x}$-direction, and the other two orthogonal directions are defined as $\mathit{y}$ and $\mathit{z}$. The origin is placed at the center of the front surface of the rotating object. The solid is rotating around the $\mathit{x}$-axis with a constant angular speed $\omega_x$. A fixed dimensionless velocity is imposed at the inlet, and a zero-gradient condition for velocity with a constant zero pressure is set at the outlet. On the surface of the solid object shown in blue in Fig. 1, a no-slip condition is employed. Throughout this study, physical quantities are normalized by the inflow velocity $u_{\mathrm{in}}^\#$ and the channel height $L^\#$, and the Reynolds number is defined as $Re = u_{\mathrm{in}}^\# L^\# / \nu^\#$ where $\nu^\#$ is the fluid kinematic viscosity. Here, the superscript \# denotes a dimensional quantity. We also define the rotating Reynolds number as $Re_\omega = 2 \omega_x^\# r_\omega^\# / \nu^\#$, where $\omega_x^\#$  and $r_\omega^\#$ are the rotating speed and the maximum radius of the rotating object, respectively.

\subsection{Grid systems for MRF and SLM}
\label{subsec2_2}
\begin{figure}[H]
\centering
\begin{subfigure}{0.48\linewidth}
    \caption{}
    \includegraphics[width=\linewidth]{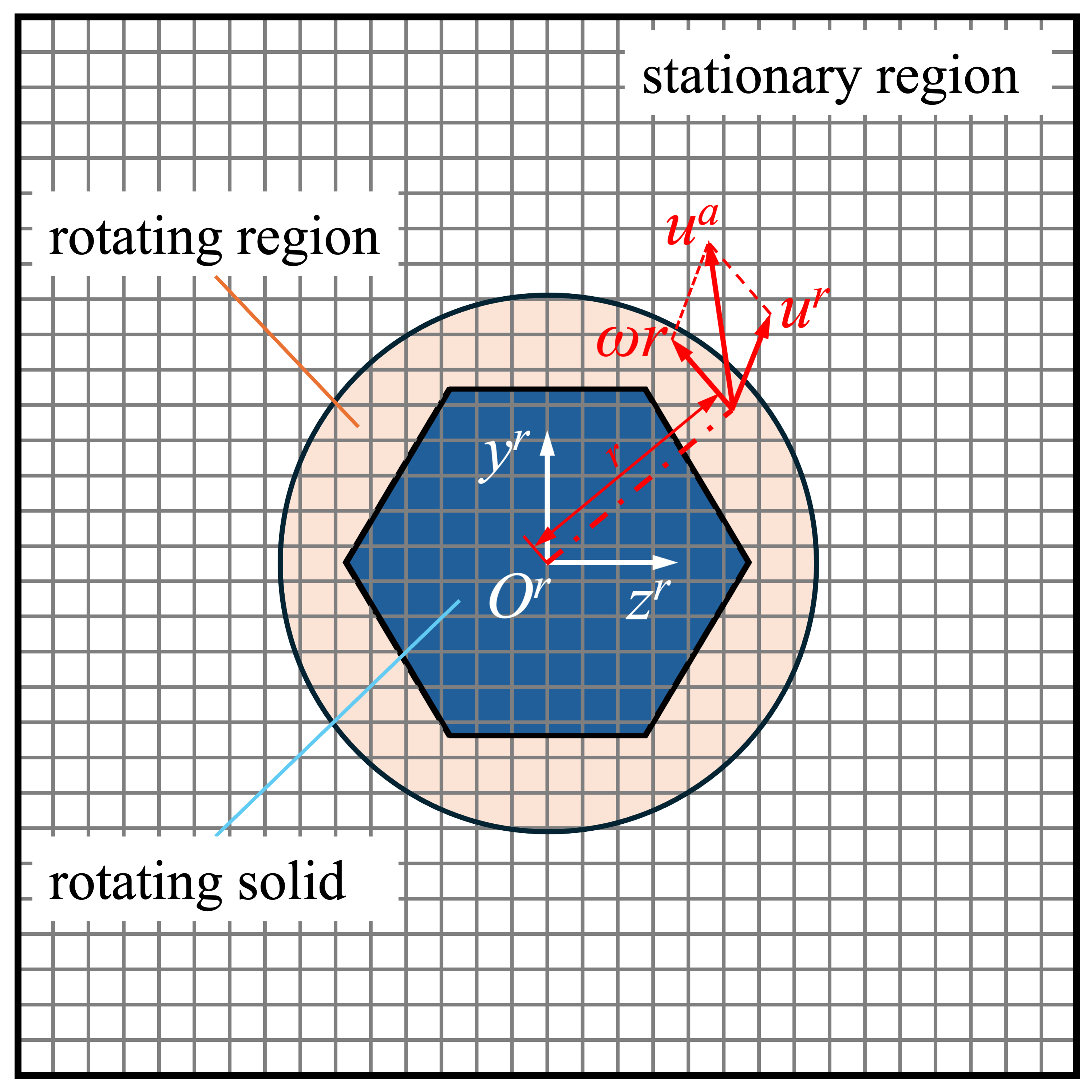}
    \centering
\end{subfigure}
\hfill
\begin{subfigure}{0.48\linewidth}
    \caption{}
    \includegraphics[width=\linewidth]{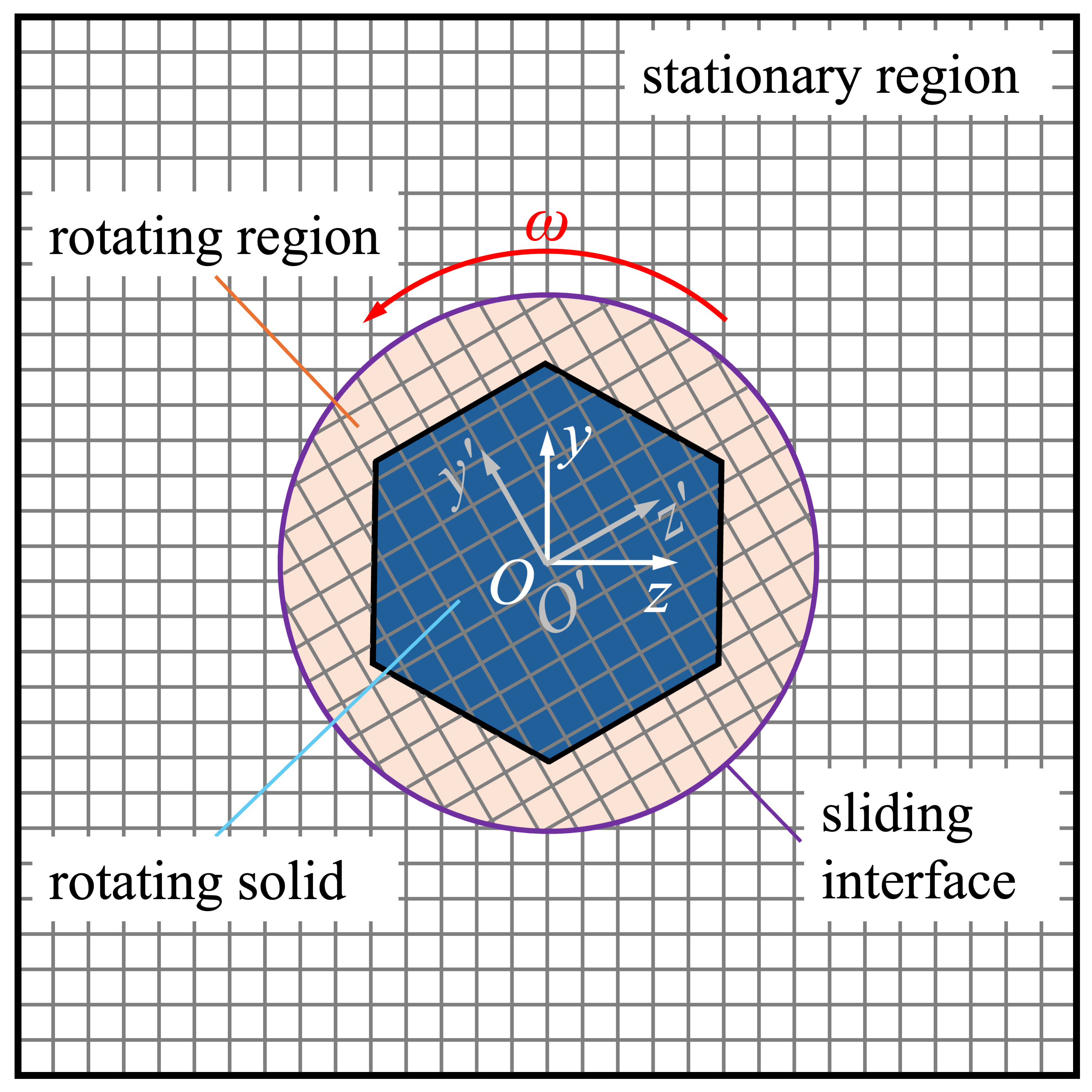}
    \centering
\end{subfigure}

\caption{Schematic of the grid systems used in (a) MRF and (b) SLM.}
\label{fig2}
\end{figure}

In order to simulate a flow around a rotating solid, both MRF and SLM approaches are considered and combined with VPM in the present study. The schematic of grid systems used for these two methods is illustrated in Fig. \ref{fig2}. In both methods, the entire computational domain is divided into a rotating region containing a rotating object with a constant angular velocity and an external stationary region.

The rotating solid object is assumed to be fixed inside the rotating region for the MRF (see Fig. \ref{fig2}(a)). In this case, only the absolute coordinate $y^r O^r z^r$ (see Fig. \ref{fig2}(a)) is used since the grids are stationary even in the rotating region. Although the coordinate does not rotate in the framework of MRF, the absolute velocity $\boldsymbol{u}^a$ inside the rotating region can be virtually decomposed into a rotating velocity $\boldsymbol{u}_\omega$  and a relative velocity $\boldsymbol{u}^r$ as $\boldsymbol{u}^a$ = $\boldsymbol{u}_\omega$ + $\boldsymbol{u}^r$ where $\boldsymbol{u}_\omega$ = $\boldsymbol{\omega} \times \boldsymbol{r}$. Here, $\boldsymbol{\omega}$ and $\boldsymbol{r}$ are the rotation speed vector and the relative location vector from the center of the rotation, respectively. By comparison, only absolute velocity is considered in the stationary region outside the rotating region.
In terms of SLM, two sets of the coordinate system $yOz$ and $y'O'z'$ (see Fig. \ref{fig2}(b)) are defined for stationary and rotating regions, respectively. The grids inside the rotating region rotate with a predefined rotation speed $\omega_x$. The interface between the interior rotating grids and the external stationary grids is defined as a sliding interface (see, the purple circle in Fig. \ref{fig2}(b)). The grid nodes on the sliding interface in the rotating and stationary regions are remapped at each step, and the stationary and rotating regions are coupled across the sliding interface through the continuity of the local velocity and stresses. Therefore, these two regions are geometrically separated but numerically coupled across the sliding interface.

\subsection{Level-set function}
\label{subsec2_3}
For both MRF and SLM, we propose to apply IBM to embed a rotating object with arbitrary geometry in the Cartesian grids of the rotating region. This is achieved by adopting the level-set function. The level-set function $\phi_0$ is defined as a signed distance function from the interface between the fluid and solid regions. Accordingly, its value is set to be zero on the interface, negative in the fluid region, and positive in the solid region. The level-set function can be converted into the phase indicator, namely, $\phi$ = 0 in the fluid region, 1 in the solid region, and 0 < $\phi$ < 1 in the interfacial region using the following formula:
\begin{equation}
\phi =
\begin{cases}
0, & \phi_0 < -\delta, \\[6pt]
\dfrac{1}{2}\sin\!\left( \dfrac{\phi_0}{\delta}\,\dfrac{\pi}{2} \right) + \dfrac{1}{2},
& -\delta \le \phi_0 \le \delta, \\[10pt]
1, & \phi_0 > \delta ,
\end{cases}
\end{equation}
where $\delta$ denotes the half-width of an interfacial region (0 < $\phi$ < 1) between the fluid and the solid regions. In the present study, the interfacial thickness $\delta$ is determined by the local diagonal grid spacing multiplied by a constant coefficient $K_\delta$. Specifically, $K_\delta$ is set as 1.5 in the present simulations. Therefore, under a fixed coefficient of $K_\delta$, the interfacial thickness becomes narrower as the grid size reduces, and the resultant numerical result should converge to the true solution. Figure 3 shows typical profiles of the level-set function $\phi_0$  and the phase indicator $\phi$  in the vicinity of the fluid-solid interface. More details on implementing the level-set function can be found in Chen et al \cite{Chen23}.

\begin{figure}[H]
\centering
\includegraphics[width=0.8\textwidth]{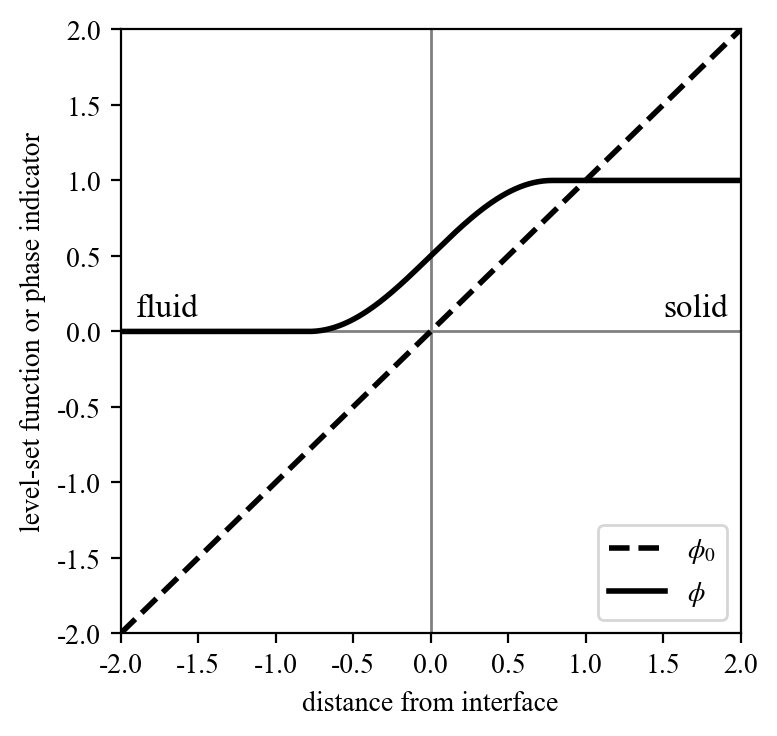}
\caption{Profiles of the level-set function $\phi_{0}$ and the phase indicator $\phi$ in the vicinity of the fluid-solid interface.}\label{fig3}
\end{figure}

\subsection{Multiple reference frame}
\label{subsec2_4}
In the MRF framework, the momentum equation of the relative velocity $u_i^r$ in the rotating coordinate system $x_i^r$  (see Fig. \ref{fig2}(a)) is written as follows \cite{Kageyama06}\cite{Combrinck17},
\begin{equation}
\frac{\partial u_i^{r}}{\partial t}
+ u_j^{r} \frac{\partial u_i^{r}}{\partial x_j^{r}}
=
- \frac{\partial p}{\partial x_i^{r}}
+ \frac{1}{Re} \frac{\partial^{2} u_i^{r}}{\partial x_j^{r} \partial x_j^{r}}
- \eta\, \phi \, u_i^{r}
- 2 \psi \, \varepsilon_{ijk} \, \omega_j \, u_k^{r}
- \psi \, \varepsilon_{ijk} \, \varepsilon_{kmn} \, \omega_j \, \omega_m \, r_n ,
\label{eq2}
\end{equation}
where $\psi$ is the space-dependent binary indicator of the rotating region, which is unity in the rotating region and zero in the stationary region. Therefore, the relative velocity $u_i^r$  is defined as
\begin{equation}
u_i^{r}
=
u_i^{a}
- \psi \, \varepsilon_{ijk} \, \omega_j \, r_k^{r} ,
\label{eq3}
\end{equation}
where the superscript $r$ and $a$ denote relative and absolute quantities, respectively. In the RHS of Eq. (\ref{eq2}), the third term, $- \eta \phi u_i^{r}$, is the artificial body forcing term to force the relative velocity to approach zero inside the rotating solid. The fourth and fifth terms, $-2 \psi \varepsilon_{ijk} \omega_j u_k^{r}$ and $- \psi \varepsilon_{ijk} \varepsilon_{kmn} \omega_j \omega_m r_n$, represent the Coriolis and centrifugal forces arising from the rotation of the reference frame, respectively.

It is common to solve the governing equation of the relative velocity, while it causes the jump of the relative velocity across the interface between rotating and stationary regions. This results in numerical instabilities as pointed out in \cite{Zhou17}\cite{Caze24}. Therefore, we solve the following governing equations for the absolute velocity components by substituting Eq. (\ref{eq3}) into Eq. (\ref{eq2}):
\begin{equation}
\frac{\partial u_i^{a}}{\partial t}
+ u_j^{r} \frac{\partial u_i^{a}}{\partial x_j^{r}}
=
- \frac{\partial p}{\partial x_i^{r}}
+ \frac{1}{Re} \frac{\partial^{2} u_i^{a}}{\partial x_j^{r} \partial x_j^{r}}
- \eta \, \phi \left( u_i^{} - \psi \, \varepsilon_{ijk} \, \omega_j \, r_k^{r} \right)
- \psi \, \varepsilon_{ijk} \, \omega_j \, u_k^{a} .
\label{eq4}
\end{equation}
The continuity equation for the absolute velocity components is also transformed into the following form:
\begin{equation}
\frac{\partial u_i^{a}}{\partial x_i^{r}}
=
\frac{\partial \bigl( \psi \, \varepsilon_{ijk} \, \omega_j \, r_k^{r} \bigr)}{\partial x_i^{r}} .
\label{eq5}
\end{equation}
By solving Eqs. (\ref{eq4})-(\ref{eq5}), the flow fields can be solved in a unified manner without introducing the velocity jump at the interface between the rotating and stationary regions. Otherwise, the jump of the relative velocity across the stationary-rotating interface must be properly treated \cite{Zhou17}\cite{Caze24}.

\subsection{Sliding mesh}
\label{subsec2_5}
When the sliding mesh technique is adopted, the geometry of the rotating object becomes stationary in the reference frame rotating with the same angular velocity. Then, the momentum equation for the absolute velocity in the absolute coordinate system is written as
\begin{equation}
\frac{\partial u_i^{a}}{\partial t}
+ u_j^{a} \frac{\partial u_i^{a}}{\partial x_j}
=
- \frac{\partial p}{\partial x_i}
+ \frac{1}{Re} \frac{\partial^{2} u_i^{a}}{\partial x_j \partial x_j}
- \eta \, \phi \left( u_i^{a} - \psi \, \varepsilon_{ijk} \, \omega_j \, r_k(t) \right) .
\label{eq6}
\end{equation}
The last term on the RHS of Eq. (\ref{eq6}), $- \eta \phi \left(u_i^{a} - \psi \varepsilon_{ijk} \omega_j r_k(t) \right)$, is the VPM term introduced to make the local velocity approach the solid rotating velocity. We noted here that, as the grids inside the rotating region rotate, the radius vector $r_i$ defined based on the absolute coordinate system should update accordingly in each grid. Then, by solving Eq. (\ref{eq6}), the flow fields within both the stationary and rotating regions, as well as the fluid and solid regions, are simulated through a unified governing equation.

Since the entire domain is separated into the rotating and stationary regions, the following boundary conditions for the absolute velocity and the stresses are defined on the sliding interface so as to couple flow quantities across these two regions,
\begin{equation}
u_i^{(a,\mathrm{ro})} \, n_i
=
u_i^{(a,\mathrm{st})} \, n_i ,
\label{eq7}
\end{equation}
\begin{equation}
\left[
- p^{\mathrm{ro}} \delta_{ij}
+ \frac{1}{Re}
\left(
\frac{\partial u_i^{(a,\mathrm{ro})}}{\partial x_j}
+
\frac{\partial u_j^{(a,\mathrm{ro})}}{\partial x_i}
\right)
\right] n_i
=
\left[
- p^{\mathrm{st}} \delta_{ij}
+ \frac{1}{Re}
\left(
\frac{\partial u_i^{(a,\mathrm{st})}}{\partial x_j}
+
\frac{\partial u_j^{(a,\mathrm{st})}}{\partial x_i}
\right)
\right] n_i ,
\label{eq8}
\end{equation}
where the superscripts ro and st denote the rotating and stationary regions, respectively. $n_i$ represents the unit vector normal to the sliding interface pointed from the rotating region towards the stationary region.
It should be noted that a pressure jump could occur across the sliding interface when solving Eq. (\ref{eq6}) directly, as reported in previous immersed boundary simulations \cite{Menez23}\cite{Wu25}. This is attributed to the inconsistency between the intermediate velocity and pressure in a conventional Pressure-Implicit with Splitting of Operators (PISO) loop. In the momentum predictor of a PISO loop, an intermediate velocity is calculated using the pressure from the last time step, while the update of the pressure due to the rotation of the rotating region is not taken into account. To solve this issue, the following pressure correction algorithm \cite{Horng18} is introduced and implemented in addition to a conventional PISO loop. The key process is to solve the following Poisson equation to determine the pressure correction function $\varphi$,
\begin{equation}
\frac{\partial^{2} \varphi}{\partial x_i \partial x_i}
=
\frac{1}{\Delta t}
\frac{\partial u_i^{*}}{\partial x_i}
\label{eq9}
\end{equation}
where $u_i^*$ is the velocity from the momentum predictor in the PISO loop. Then, the intermediate velocity $u_i^**$  and the intermediate pressure $p^**$  are updated as follows,
\begin{equation}
u_i^{**}
=
u_i^{*}
-
\Delta t \, \frac{\partial \phi}{\partial x_i}
\label{eq10}
\end{equation}
\begin{equation}
p^{**} = p^{*} + \varphi
\label{eq11}
\end{equation}
and these are used for the momentum corrector in the PISO loop.

\subsection{Numerical setup}
\label{subsec2_6}
So far, we have derived the governing equations based on VPM combined with MRF and SLM, respectively. These equations are numerically solved through the finite volume method (FVM), and they are implemented into OpenFOAM (version 8), an open-source CFD code. The advection and diffusion terms are discretized using second-order linear scheme, and the transient term is discretized using first-order implicit Euler scheme. Note that for SLM, the grids in the rotating and stationary regions are coupled through the technique of arbitrary mesh interface (AMI) in OpenFOAM. This operates by projecting grid nodes of the rotating region into the stationary region across the sliding interface.

\section{Verification cases}
\label{sec3}
\subsection{Case settings}
\label{subsec3_1}
\begin{figure}[H]
\centering
\includegraphics[width=0.8\textwidth]{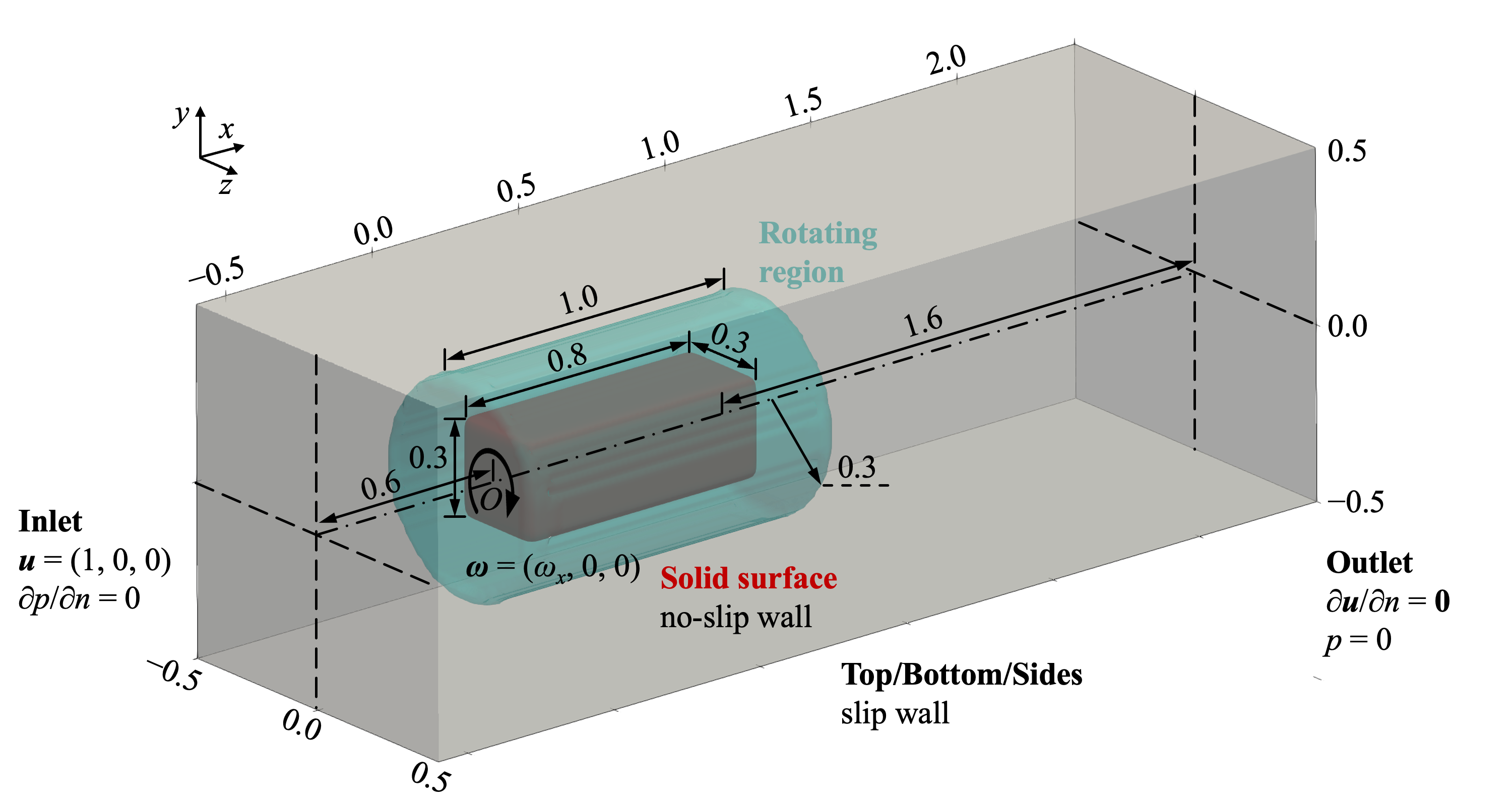}
\caption{Case setting of the flow around a rotating cuboid.}\label{fig4}
\end{figure}

An incompressible laminar flow past a rotating cuboid inside a square channel is considered in the present study, where the inflow velocity and the channel height are taken as the characteristic velocity and length scales, respectively. The origin is defined at the center of the front surface of the rotating cuboid. As illustrated in Fig. 4, the computational domain with a size of [-0.6, 2.4]$\times$[-0.5, 0.5]$\times$[-0.5, 0.5] is defined. A rotating cuboid with a streamwise length of 0.8 and a side length of 0.3 is located 0.6 away from the inlet. The rotating axis of the solid aligns with the central axis in the $x$-direction of $y$ = $z$ = 0. The rotating region is defined as a cylindrical region with a radius of 0.3 from $x$ = -0.1 to x = 0.9 as shown in the cyan circular cylinder surface in Fig. \ref{fig4}, so that it covers the rotating cuboid. A fixed uniform inflow is imposed at the inlet, and a zero gradient velocity condition is set at the outlet. A no-slip condition is imposed on the surfaces of the rotating object. Free-slip conditions are set on the side boundaries to minimize their effects. In the present study, two sets of Reynolds numbers are defined based on the inflow condition and the rotating condition, respectively. The Reynolds number Re based on the inflow velocity and the channel height is fixed as 111.11. The rotating speeds systematically increased from 2, 4, 6, 8 to 10, resulting in the rotating Reynolds number $Re_\omega$ changing from 20, 40, 60, 80, to 100.

\subsection{Numerical settings}
\label{subsec3_2}
\begin{figure}[htbp]
  \centering

  \begin{subfigure}{\linewidth}
    \centering
    \caption{}
    \includegraphics[width=0.95\linewidth]{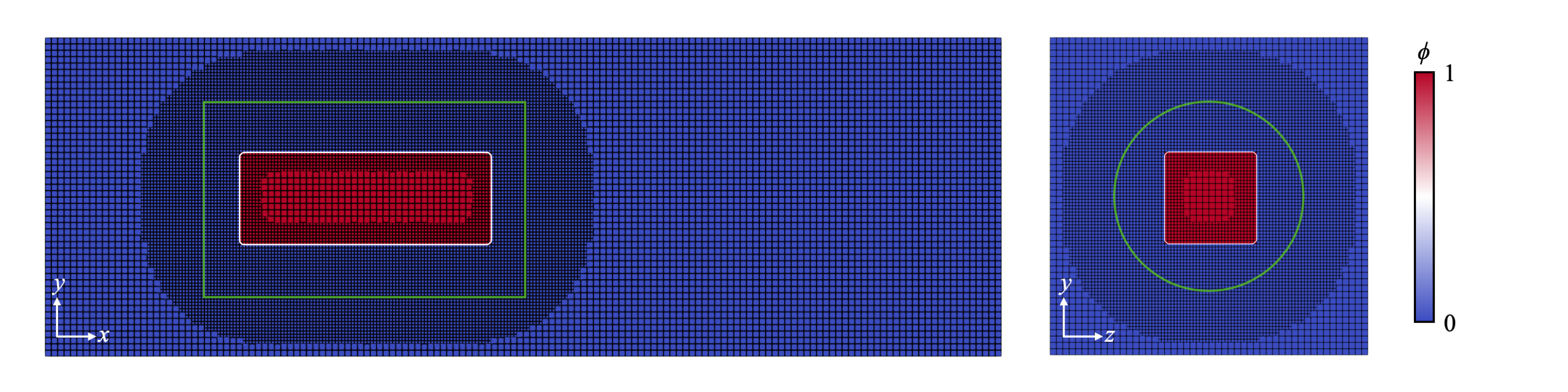}
  \end{subfigure}\par\medskip

  \begin{subfigure}{\linewidth}
    \centering
    \caption{}
    \includegraphics[width=0.95\linewidth]{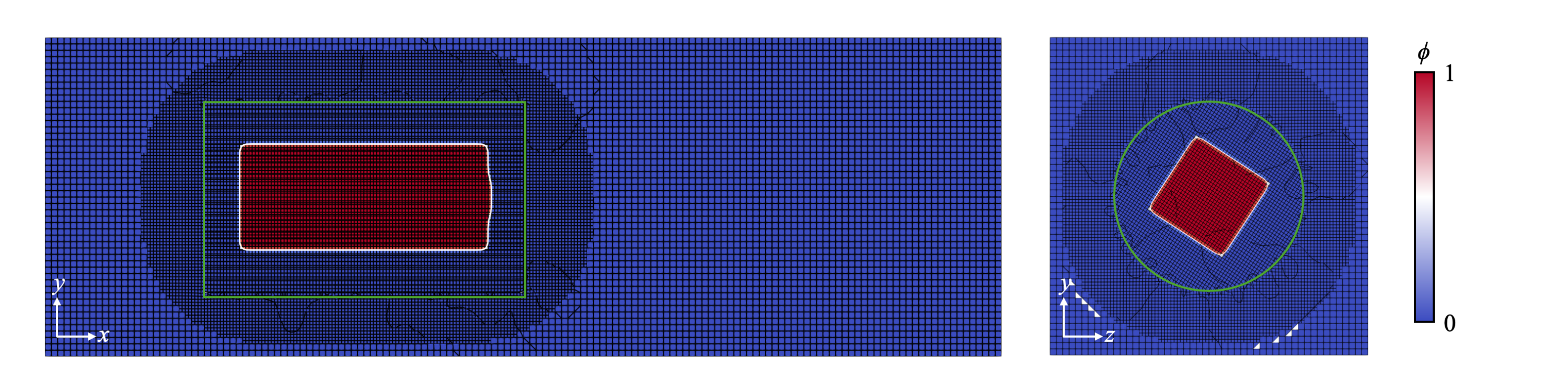}
  \end{subfigure}\par\medskip

  \begin{subfigure}{\linewidth}
    \centering
    \caption{}
    \includegraphics[width=0.95\linewidth]{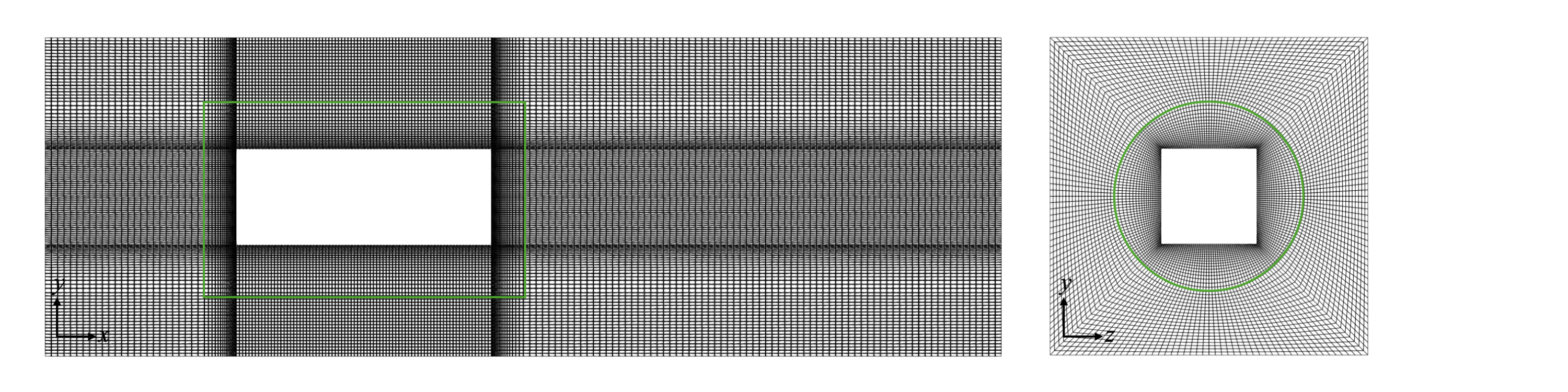}
  \end{subfigure}\par\medskip

  \begin{subfigure}{\linewidth}
    \centering
    \caption{}
    \includegraphics[width=0.95\linewidth]{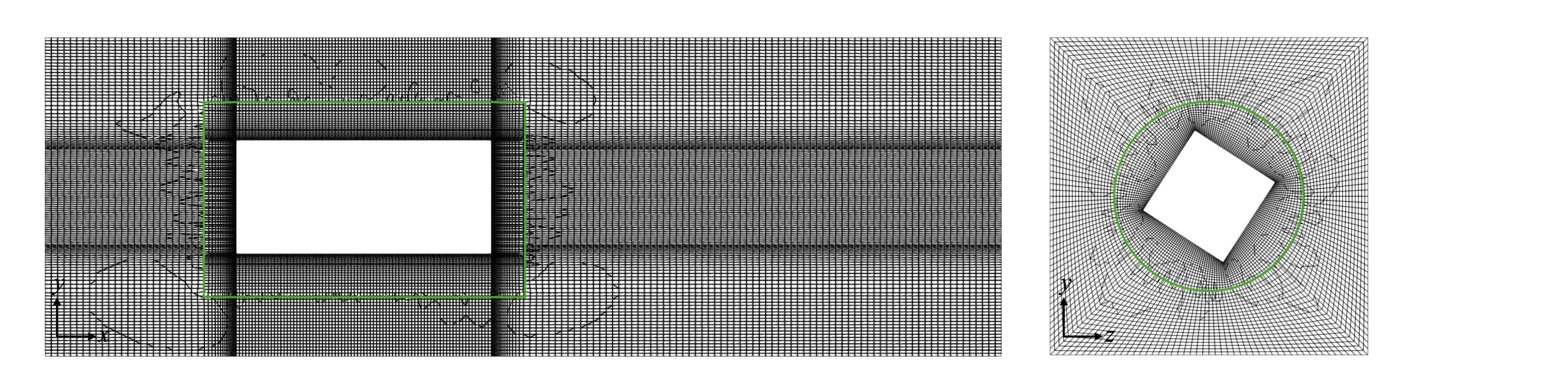}
  \end{subfigure}

  \caption{Distribution of Cartesian grids for (a) VPM with MRF and (b) VPM with SLM and structured grids for (c) BFM with MRF and (d) BFM with SLM. Here, the white and the green line indicate the fluid-solid interface and static-rotating interface, respectively.}
  \label{fig5}
\end{figure}

For the present VPM simulations, the Cartesian grids are generated for VPM with MRF (see Fig. \ref{fig5}(a)) and VPM with SLM (see Fig. \ref{fig5}(b)). For their validation, the corresponding simulations based on the body-fitted method (BFM) with MRF and SLM are also performed using the standard OpenFOAM solvers, simpleFoam and pimpleFoam, respectively. The body-fitted grids are also generated for BFM with MRF (see Fig. \ref{fig5}(c)) and BFM with SLM (see Fig. \ref{fig5}(d)). In unsteady simulations using SLM, an adaptive timestep is adopted to ensure that the maximum Courant number is less than 0.3. Various studies have been conducted to perform body-fitted simulations on rotating machinery using OpenFOAM based on MRF \cite{Huang19} and SLM \cite{Madina21}\cite{Hundshagen20}. The SLM-based simulation results show good agreement with experimental measurement in both performance \cite{Madina21}\cite{Hundshagen20} and flow fields \cite{Schoberiri13}. Specifically, the relative error of the predicted pump head is reported to be less than 3\% in Hundshagen \cite{Hundshagen20}. We first validate the present VPM by comparing the corresponding BFM results for each framework of MRF and SLM. Then, we further discuss the differences between simulation results obtained by MRF and SLM.

\subsection{Grid convergence study}
\label{subsec3_3}
First, the grid convergence studies are performed for both BFM and VPM under the rotating condition $Re_\omega$ = 100. For the simulations with body-fitted grids, the number of grid points is systematically increased from 0.3 million to 7.1 million. As listed in Table 1, when the number of grid points exceeds 2 million, the relative deviations of both MRF and SLM compared to the corresponding results using the finest grids are less than 0.5\%. Therefore, the body-fitted mesh with 2.1 million grid points is used in the following study. Table 2 summarizes the grid convergence results of the present VPM for both MRF and SLM. We systematically increase the number of grid points in the $x$, $y$, $z$-directions, as ($N_x \times N_y \times N_z$ = (75$\times$25$\times$25), (120$\times$40$\times$40), (150$\times$50$\times$50), (200$\times$67$\times$67), (300$\times$100$\times$100), resulting the minimum grid sizes of 0.020, 0.013, 0.010, 0.008, and 0.005, respectively, after applying first-level adaptive mesh refinement (AMR) based on the local value of the level-set function. Taking the simulation result using the finest grid as the reference value, the relative deviation also reduces to 0.5\% for both MRF and SLM. Therefore, the Cartesian grids of $N_x \times N_y \times N_z$ = 150$\times$50$\times$50 are adopted in the following VPM simulations. Using the above-mentioned grids, the steady simulations of 10,000 timesteps using VPM with MRF and BFM with MRF take 51 and 43 minutes, respectively, for parallel computations with 24 CPUs (AMD EPYC 7702). As for SLM, the unsteady simulations for dimensionless time period $T$ = 10 using VPM with SLM and BFM with SLM take 1,319 and 4,525 minutes, respectively, with 768 processors parallel computation on the supercomputer Fugaku. This suggests the great potential of reducing the computational cost by the present VPM.

\begin{table}[H]
\centering
\caption{Convergence check of BFM.}
\label{tab1}
\resizebox{\linewidth}{!}{%
\begin{tabular}{c cc cc}
\toprule
\makecell{\textbf{Number of grid}\\ \textbf{points}}
& \multicolumn{2}{c}{\textbf{MRF}}
& \multicolumn{2}{c}{\textbf{SLM}} \\
\cmidrule(lr){2-3}\cmidrule(lr){4-5}
& \makecell{\textbf{Pressure drop}}
& \makecell{\textbf{Relative}\\ \textbf{deviation}}
& \makecell{\textbf{Pressure drop}}
& \makecell{\textbf{Relative}\\ \textbf{deviation}} \\
\midrule
292,625   & 0.361280 & $-0.68\%$ & 0.377378 & $-0.80\%$ \\
904,020   & 0.361600 & $-0.60\%$ & 0.385677 & 1.38\% \\
2,110,000 & 0.362612 & $-0.32\%$ & 0.379140 & $-0.34\%$ \\
4,159,575 & 0.363019 & $-0.21\%$ & 0.380363 & $-0.01\%$ \\
7,097,625 & 0.363771 & 0.00\%    & 0.380416 & 0.00\% \\
\bottomrule
\end{tabular}
}
\end{table}

\begin{table}[htbp]
\centering
\caption{Convergence check of VPM.}
\label{tab2}
\resizebox{\linewidth}{!}{%
\begin{tabular}{c c c cc cc}
\toprule
$\mathit{N_x \times N_y \times N_z}$ &
\makecell{\textbf{Minimum}\\ \textbf{grid size}\\ \textbf{after AMR}} &
\makecell{\textbf{Number of}\\ \textbf{grid points}\\ \textbf{after AMR}} &
\multicolumn{2}{c}{\textbf{MRF}} &
\multicolumn{2}{c}{\textbf{SLM}} \\
\cmidrule(lr){4-5}\cmidrule(lr){6-7}
& & &
\makecell{\textbf{Pressure}\\ \textbf{drop}} &
\makecell{\textbf{Relative}\\ \textbf{deviation}} &
\makecell{\textbf{Pressure}\\ \textbf{drop}} &
\makecell{\textbf{Relative}\\ \textbf{deviation}} \\
\midrule
75$\times$25$\times$25     & 0.020 & 149,285   & 0.373117 & 2.12\%     & 0.378913 & $-2.95\%$ \\
120$\times$40$\times$40    & 0.013 & 445,918   & 0.380970 & 4.27\%     & 0.378993 & $-2.93\%$ \\
150$\times$50$\times$50    & 0.010 & 717,510   & 0.363356 & $-0.55\%$  & 0.392461 & 0.52\% \\
200$\times$67$\times$67    & 0.008 & 1,523,131 & 0.365479 & 0.03\%     & 0.391838 & 0.36\% \\
300$\times$100$\times$100  & 0.005 & 4,345,981 & 0.365373 & 0.00\%     & 0.390434 & 0.00\% \\
\bottomrule
\end{tabular}
}
\end{table}

\subsection{Results and discussions}
\label{subsec3_4}
In order to access the accuracy of the present VPM, two commonly-used performance indices for rotating machinery, namely, the pressure drop from the inlet to the outlet $\Delta p$, and the torque acting on the rotating solid $T_i$ are calculated for all these schemes for comparison. For VPM, the torque is calculated as
\begin{equation}
T_i
=
\iiint_{\Omega}
\varepsilon_{ijk} \, r_j \, \eta \, \phi
\left(
u_k
- \psi \, \varepsilon_{kmn} \, \omega_m \, r_n
\right)
\, dV
\label{eq12}
\end{equation}
where $\Omega$ denotes the entire domain. Note that the VPM term $- \eta \phi \left(u_i^{a} - \psi \varepsilon_{ijk} \omega_j r_k(t) \right)$) is the local force arising from the rotating object, and the cross product of this force and the radius vector provides the torque. For BFM, the torque is calculated by the surface integration of the fluid stress $\tau_i$ acting on the fluid-solid boundary $\partial \Omega_s$ as

\begin{figure}[H]
\centering
\begin{subfigure}{0.48\linewidth}
    \caption{}
    \includegraphics[width=\linewidth]{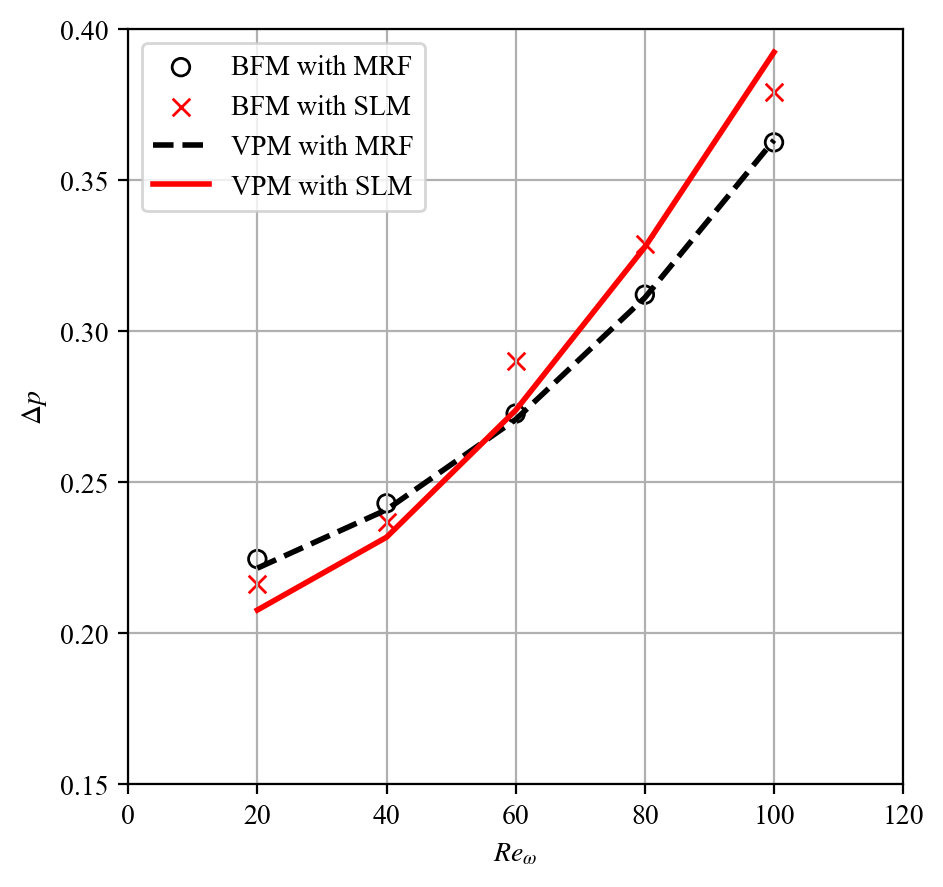}
    \centering
\end{subfigure}
\hfill
\begin{subfigure}{0.48\linewidth}
    \caption{}
    \includegraphics[width=\linewidth]{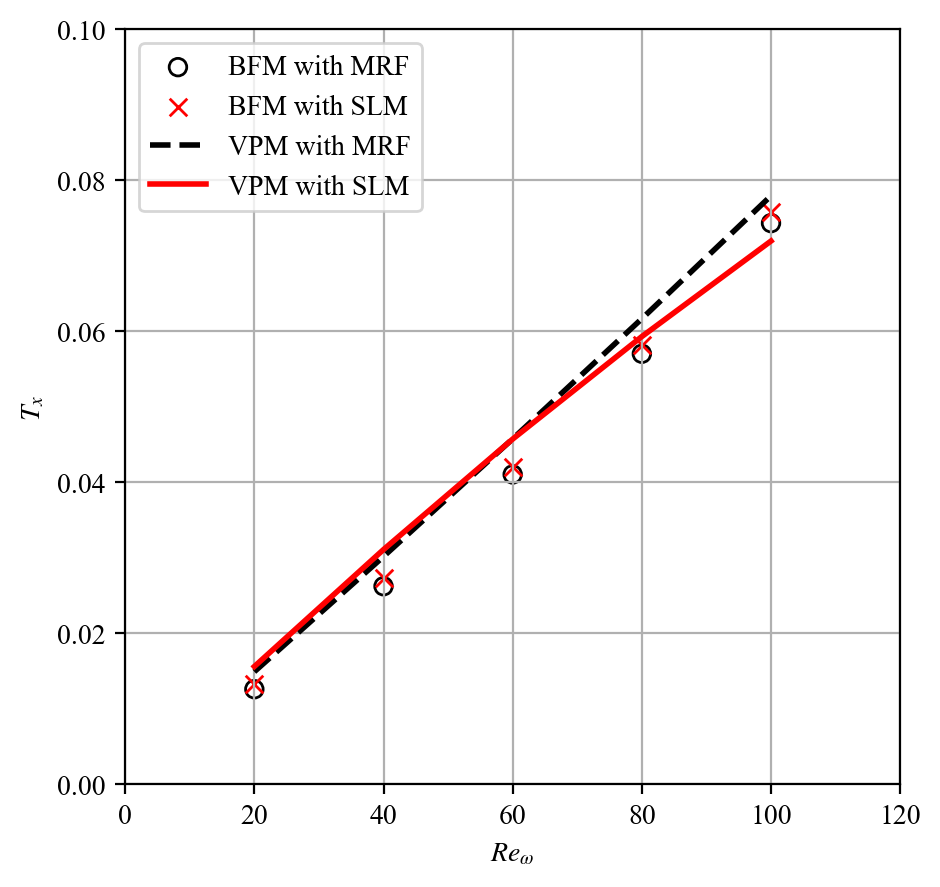}
    \centering
\end{subfigure}

\caption{Simulation results of (a) pressure drop $\Delta p$ and (b) torque $T_x$ under different $Re_\omega$ using various numerical schemes.}
\label{fig6}
\end{figure}

Figures \ref{fig6}(a) and (b) show the pressure drop $\Delta p$ and torque $T_x$, respectively, under different $Re_\omega$ obtained by the present VPM and BFM in the framework of MRF and SLM. As shown in Fig. \ref{fig6}(a), the pressure drops by VPM with MRF and BFM with MRF show reasonable agreement. By comparison, when SLM is adopted, for $Re_\omega$ < 80, $\Delta p$ predicted by VPM with SLM is underestimated compared to the value obtained by BFM with SLM due to the delayed boundary layer separation. In contrast, for $Re_\omega$ > 80, the issue of the pressure jump across the interfacial region discussed in Sec. 2.5 becomes dominant, resulting in the overestimation of the $\Delta p$ by VPM. As for $T_x$ shown in Fig. \ref{fig6}(b), the results by VPM are slightly higher than those by BFM in both the frameworks of MRF and SLM. In the present diffuse-interface VPM, the shear stress acting on the sharp solid interface in BFM is represented by a diffused VPM body forcing term within a finite-thickness interfacial region. This effectively increases the radius of the rotating object, and it might cause the overestimation of Tx.
In terms of the comparison between MRF and SLM, $\Delta p$ predicted by MRF is slightly higher than that by SLM under low $Re_\omega$ cases. The inertia sources introduced by the MRF formulation assume a fully developed rotating flow, while SLM captured the weaker swirls. This may explain the overestimate in the pressure drop for MRF at low rotation speeds. By comparison, as the rotation speed increases, $\Delta p$ predicted by MRF becomes smaller than that of the SLM. This is attributed to the fact that the transient interaction between rotating and stationary parts is not considered in MRF, so that momentum loss due to rotor-stator interaction is not well resolved. The necessity of applying SLM to produce unsteady flow characteristics is also reported in \cite{Schoberiri13} and \cite{Song24}.

For a quantitative evaluation, Table 3 summarizes the mean relative deviations under different $Re_\omega$ of the present VPM compared with the corresponding BFM. The quantitative comparison indicates that the relative deviations of the pressure drops for both VPM with MRF and VPM with SLM are less than 5\%, while the relative deviations of the torques are around 8\% and 9\% for VPM with MRF and VPM with SLM, respectively. This validates the good accuracy of the present VPM.

\begin{table}[htbp]
\centering
\caption{Mean relative deviations of the VPM compared to BFM.}
\label{tab3}
\resizebox{\linewidth}{!}{%
\begin{tabular}{l l c c}
\toprule
\textbf{VPM} & \textbf{Reference BFM} & \textbf{Pressure drop $\Delta p$} & \textbf{Torque $T_x$} \\
\midrule
VPM with MRF & BFM with MRF & 0.69\% & 11.58\% \\
VPM with MRF & BFM with SLM & 4.08\% & 8.03\% \\
VPM with SLM & BFM with SLM & 3.10\% & 9.33\% \\
\bottomrule
\end{tabular}
}
\end{table}

\begin{figure}[htbp]
  \centering

  \begin{subfigure}{\linewidth}
    \centering
    \caption{}
    \includegraphics[width=0.95\linewidth]{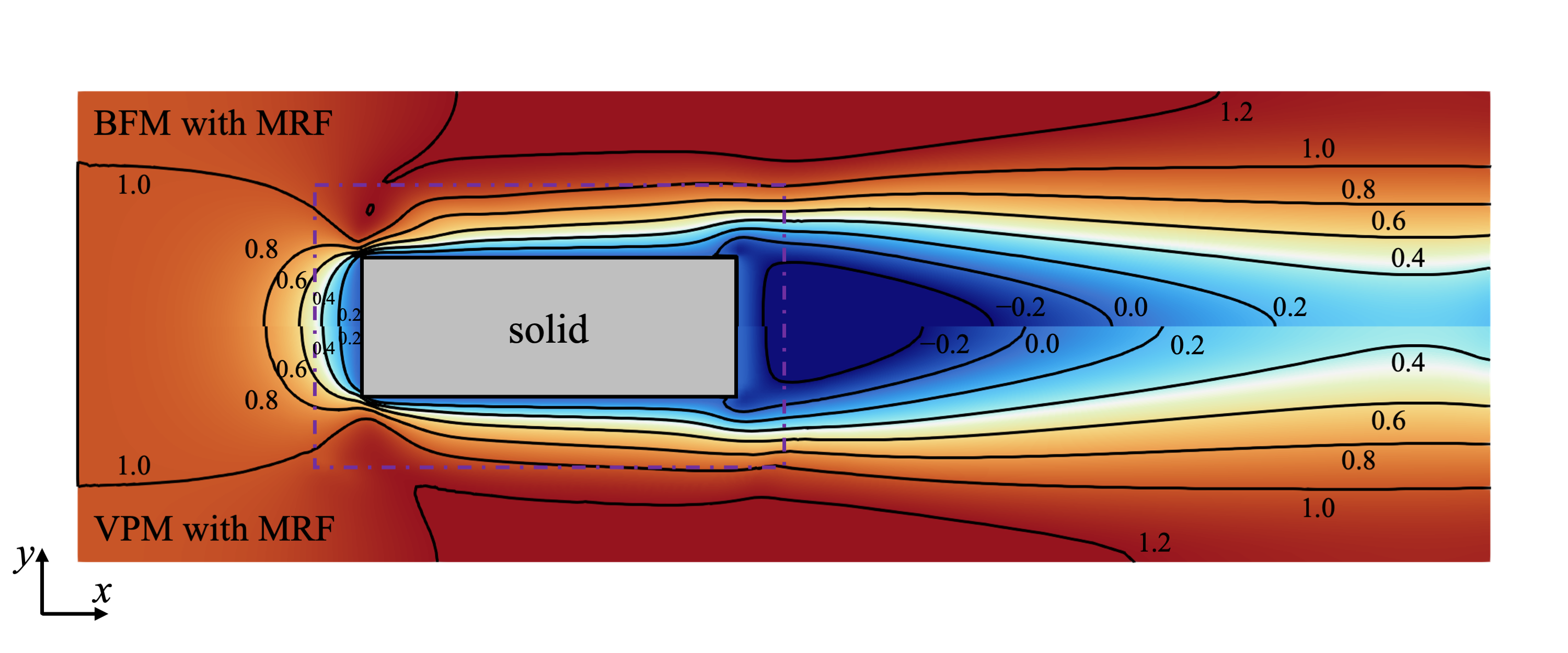}
  \end{subfigure}\par\medskip

  \begin{subfigure}{\linewidth}
    \centering
    \caption{}
    \includegraphics[width=0.95\linewidth]{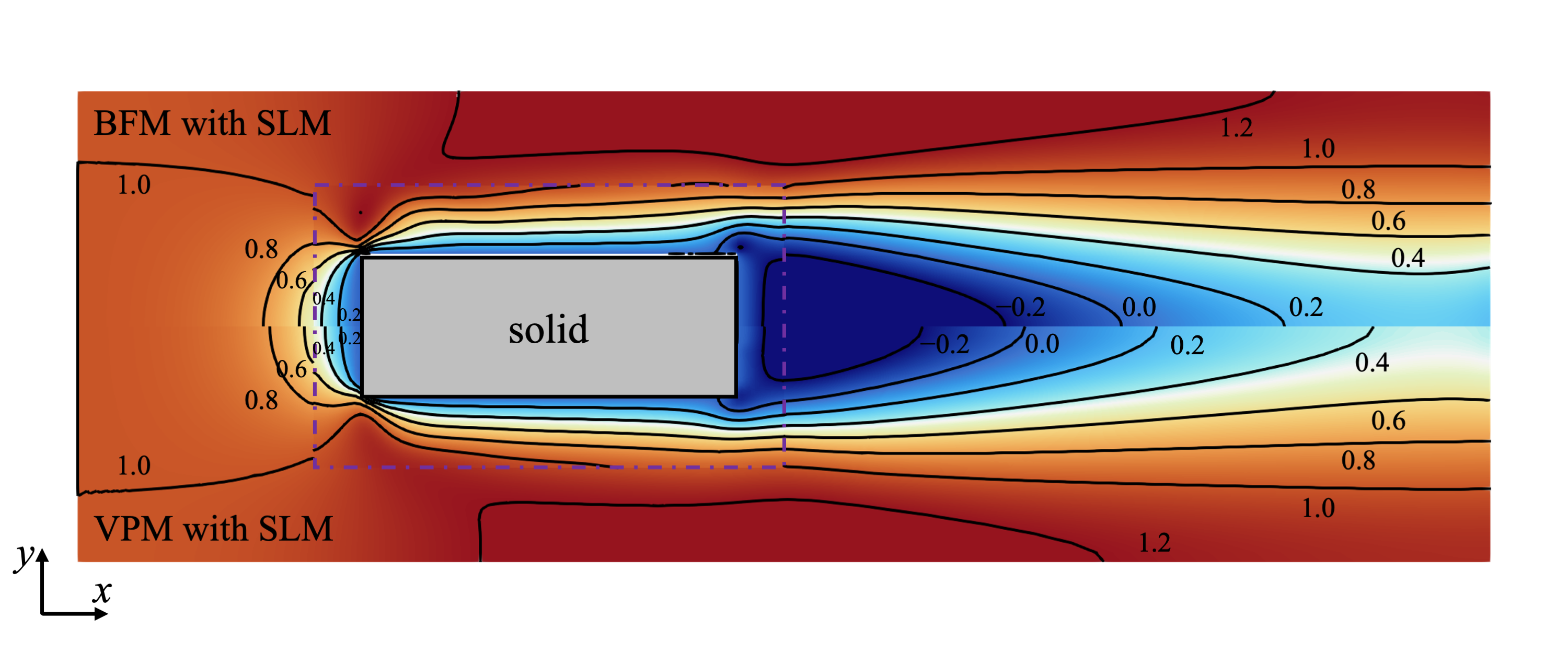}
  \end{subfigure}\par\medskip

  \begin{subfigure}{\linewidth}
    \centering
    \includegraphics[width=0.5\linewidth]{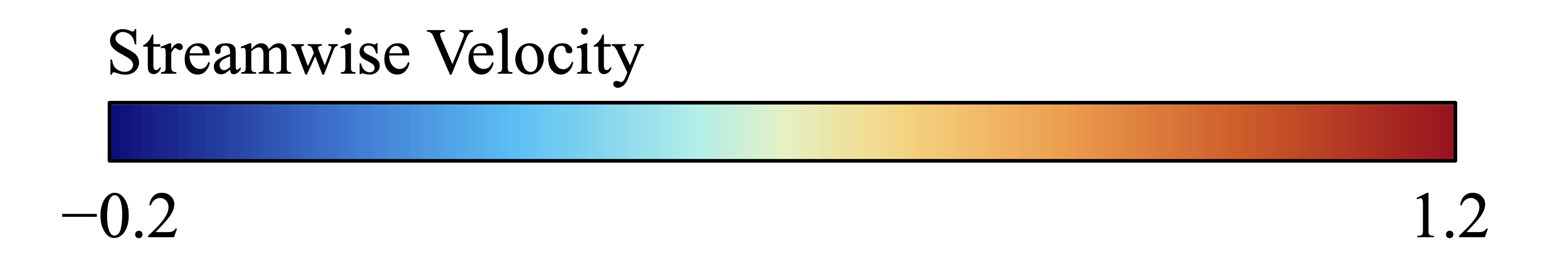}
  \end{subfigure}\par\medskip

  \caption{Distribution of streamwise velocity on the plane $z$ = 0 for (a) MRF and (b) SLM.}
  \label{fig8}
\end{figure}

\begin{figure}[htbp]
  \centering

  \begin{subfigure}{\linewidth}
    \centering
    \caption{}
    \includegraphics[width=0.95\linewidth]{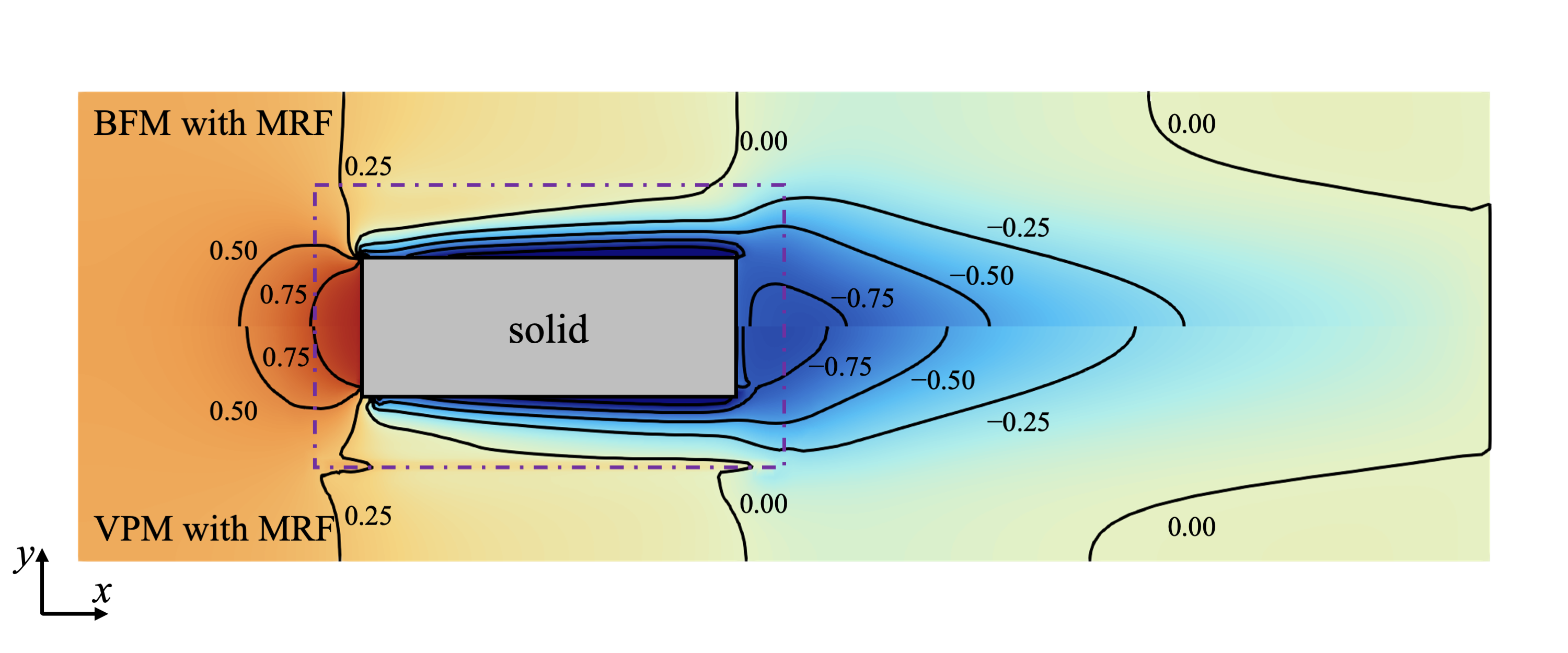}
  \end{subfigure}\par\medskip

  \begin{subfigure}{\linewidth}
    \centering
    \caption{}
    \includegraphics[width=0.95\linewidth]{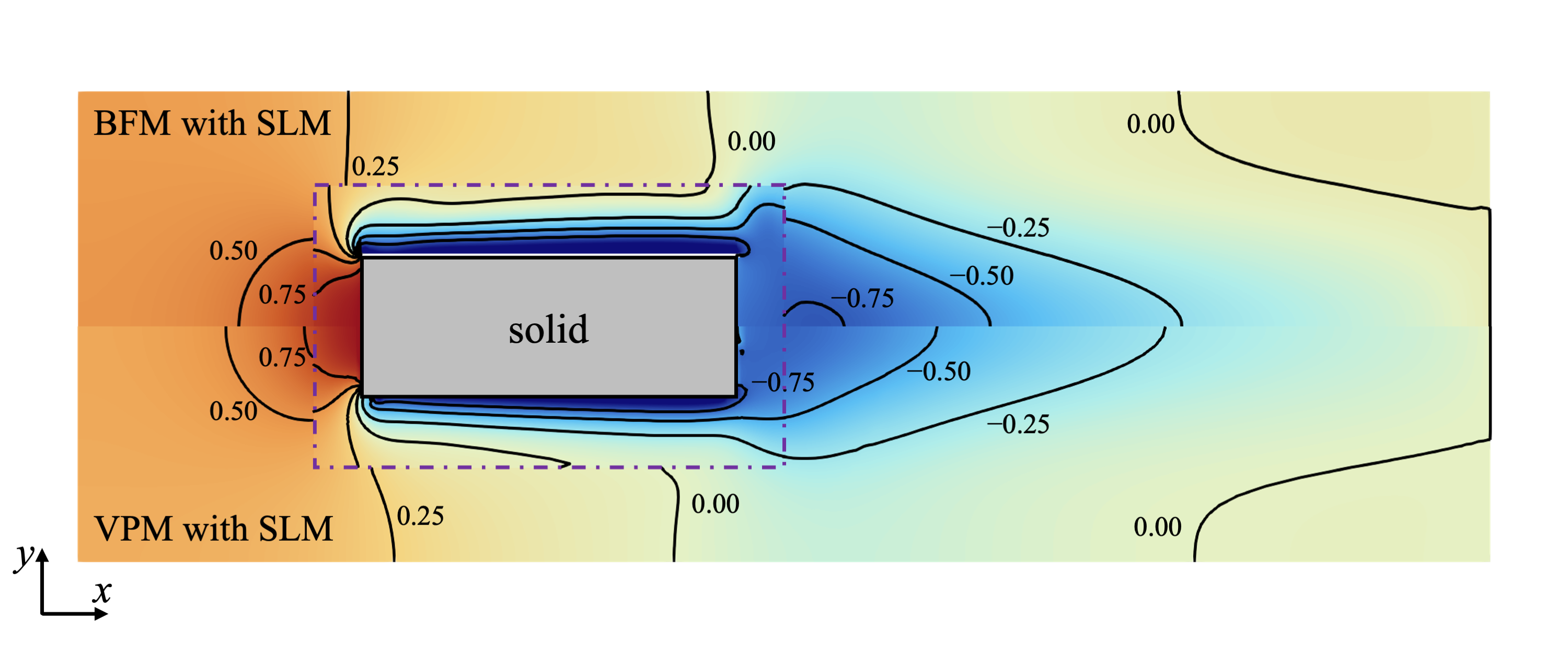}
  \end{subfigure}\par\medskip

  \begin{subfigure}{\linewidth}
    \centering
    \includegraphics[width=0.5\linewidth]{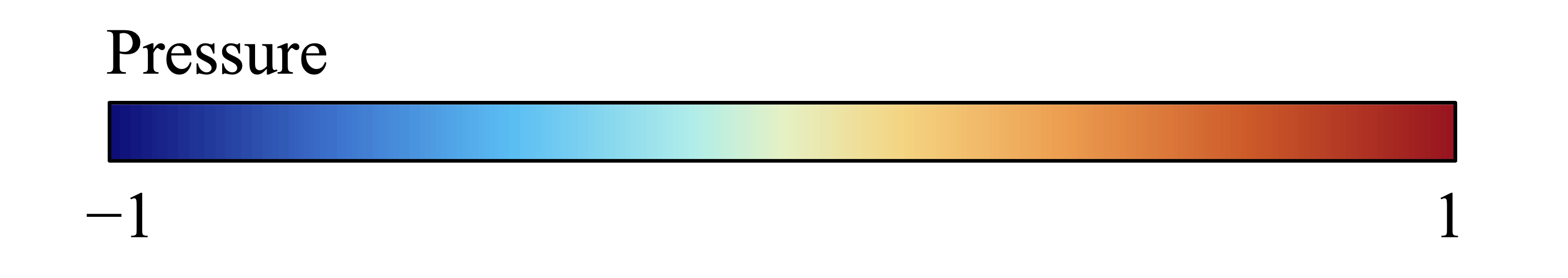}
  \end{subfigure}\par\medskip

  \caption{Distribution of pressure on the plane $z$ = 0 for (a) MRF and (b) SLM. Hereafter, the upper half and the lower half show the BFM and VPM results respectively. The purple dash dotted box indicates the interface between the rotating and stationary regions.}
  \label{fig7}
\end{figure}

To further compare the simulation results of the present VPM and BFM, Figures \ref{fig7} and \ref{fig8} show the comparison of the distributions of the pressure and the streamwise velocity on the plane z = 0 between BFM and the present VPM. The VPM results generally show good agreement with the BFM results in both pressure and velocity fields. We noted here that non-negligible discontinuity can be confirmed for both BFM and VPM, especially in the pressure field at the interface between rotating and static regions shown by a dash-dot line (see Fig. \ref{fig7}). The good agreement in the pressure fields between the present VPM and BFM also explains the consistent prediction in the pressure drop shown in Fig. 6 and Table 3. The relatively large high-pressure region upstream of the leading edge of the rotating cuboid (see lower half in Fig. \ref{fig7}(b)) also explains the overestimation of the pressure drop by VPM with SLM (see red solid lines in Fig. \ref{fig6}(a)). In terms of the streamwise velocity fields, as shown in Fig. 8, the wake region predicted by VPM is smaller than that by BFM. This is attributed to the delayed boundary layer separation predicted by VPM, arising from the non-zero velocity on the solid surface. Even though a strong forcing is applied to achieve a no-slip condition in VPM, there always exists a small but non-zero velocity on a solid surface. Such a phenomenon is also reported in the simulation of flow past a circular cylinder using VPM by Brown-Dymkoski et al \cite{Brown14}, and it can be improved by further refining the grids around the fluid-solid interface. The discrepancy in the simulated boundary layer separation also explains the relatively large difference in the predicted torque shown in Fig. \ref{fig6} and Table \ref{tab3}, since the torque is calculated directly from the velocity fields around the solid interface.

\section{Conclusions}
\label{sec4}
In order to develop numerical schemes for simulating flows around rotating solids with complex shapes, we developed a novel volume penalization method (VPM) combined with multiple reference frame (MRF) and sliding mesh (SLM). The proposed schemes allow for embedding rotating objects with arbitrary geometries in a Cartesian grid system using the level-set function representing a fluid-solid interface.
For VPM with MRF, the governing equations based on the absolute velocities are derived with additional source terms to take into account inertia effects in a rotating frame. For VPM with SLM, the entire grids are divided into the rotating and stationary regions. These two regions are geometrically separated, but numerically coupled across the sliding interface.

The proposed VPM has been applied to flow around a rotating cuboid with different rotating Reynolds number $Re_\omega$. The simulations using the present VPM with MRF and VPM with SLM are compared with corresponding simulations with the body-fitted method (BFM), namely, BFM with MRF and BFM with SLM. The relative deviations of predicted pressure drop and torque are around 5\%, which validates the present VPM.

The major advantage of the proposed numerical schemes lies in their unique feature of solving fluid and solid regions with a unified governing equation. This feature allows for performing forward simulations without generating a body-fitted mesh. In addition, the proposed scheme can be extended to shape and topology optimization in a straightforward manner. Previous studies \cite{Kametani20}-\cite{Liu25} have validated the superior performance of optimizing flow and heat transfer problems based on VPM. Such extensions can be considered in future work.

\section*{Acknowledgments}
This work was partially supported by MEXT as "Program for Promoting Researches on the Supercomputer Fugaku" (Drastic acceleration of the industrial applications of HPC through AI and research and development of new computational methods for the next era; Grant Number JPMXP1020230321) and used computational resources of (supercomputer Fugaku provided by the RIKEN Center for Computational Science; Project ID: hp230210, hp240221, hp250234).  Y.H. gratefully acknowledges the supports from JSPS KAKENHI Grant Number JP23K26034, and also Adaptable and Seamless Technology Transfer Program through Target-driven R\&D (A-STEP) from Japan Science and Technology Agency (JST) Grant Number JPMJTR232D.

\section*{Declaration of interests}
The authors declare that they have no known competing financial interests or personal relationships that could have appeared to influence the work reported in this paper.

\end{document}